\documentclass[aps,prb,twocolumn,showpacs]{revtex4-1}
\usepackage{graphicx, color}
\usepackage[caption=false]{subfig}
\usepackage{amsmath, amsthm, amssymb}
\usepackage{empheq}
\usepackage{bm}
\usepackage{natbib}
\usepackage{hyperref}
\newcommand{\mb}[1]{\mathbf{#1}}
\newcommand{\pd}{\partial}

\newcommand{\tr}{\textrm{Tr}}
\newcommand{\up}{\uparrow}
\newcommand{\down}{\downarrow}

\def\e{\epsilon}

\def\ra{\rangle}
\def\sgn{\textrm{sgn}}
\newcommand{\comments}[1]{} 
\begin{document}
\title{Geometrical engineering of a two-band Chern insulator in two dimensions with arbitrary topological index} 

\author{Doru Sticlet}
\affiliation{Laboratoire de Physique des Solides, CNRS UMR-8502, Univ. Paris Sud, 91405 Orsay Cedex, France}
\author{Frederic Pi\'echon}
\affiliation{Laboratoire de Physique des Solides, CNRS UMR-8502, Univ. Paris Sud, 91405 Orsay Cedex, France}
\author{Jean-No\"el Fuchs}
\affiliation{Laboratoire de Physique des Solides, CNRS UMR-8502, Univ. Paris Sud, 91405 Orsay Cedex, France}
\author{Pavel Kalugin}
\affiliation{Laboratoire de Physique des Solides, CNRS UMR-8502, Univ. Paris Sud, 91405 Orsay Cedex, France}
\author{Pascal Simon}
\affiliation{Laboratoire de Physique des Solides, CNRS UMR-8502, Univ. Paris Sud, 91405 Orsay Cedex, France}
\begin{abstract}
Two-dimensional 2-band insulators breaking time reversal symmetry can present topological phases indexed by a
topological invariant called the Chern number. We propose an efficient procedure to determine this topological index, which makes possible to conceive 2-band, tight-binding Hamiltonians with arbitrary Chern numbers. The technique is illustrated by a step by step construction of a model exhibiting five topological phases indexed by Chern numbers $\{0,\pm 1\pm 2\}$.
On a finite cylindrical geometry,
this insulator possesses up to two edge states which are characterized analytically. 
The model can be combined with its time reversal copy to form a quantum spin Hall insulator. It is shown that edge states in the latter can be destroyed by a time reversal invariant one-particle perturbation if the Chern number equals $\pm 2$.

\end{abstract}
\maketitle

\section{Introduction}
\label{sec:intro}
The Chern insulators were first theoretically introduced by Haldane in 1988,\cite{Haldane} 
when it was proved that it is possible to conceive a two-dimensional (2D) 
band model supporting integer Hall effect without a net magnetic field. 
That was an insulator realizing the quantum anomalous Hall effect (QAH), with only local magnetic fields, breaking time reversal symmetry (TRS), and with bands characterized by a topological index, the Chern number. The presence of phases with a non-zero Chern number is related to the  appearance of a quantified conductance at the edges of the insulator. Although a close relative of the regular integer quantum Hall effect, which is also indexed by a similar topological number,\cite{Thouless} QAH effect proved to be quite elusive. Only recent theoretical and experimental work  suggests that QAH effect could be realized in single layer and bilayer graphene.\cite{Levitov,*Qiao,*Tse,*Jung,*Ding, *Jung,* Levitov1,*Weitz,*Martin}

The idea of a Chern insulator laid dormant, until it was realized that by combining a Chern insulator with its time reversal copy produces a quantum spin Hall (QSH) insulator.\cite{KMGraphene}
It was proposed that spin-orbit interaction in graphene gaps the system such that effectively two-copies of Haldane's Chern insulator are realized, one for a given spin orientation and its time reversed mirror copy for the opposite orientation.\cite{KMGraphene} As a result, when this system is put on a  Hall bar,  
two distinct robust counter-propagating spin-polarized edge states show up.
These time reversal invariant QSH insulators are characterized by a new $\mathbb{Z}_2$ topological number.\cite{KMZ2, *MooreZ2} While the Chern number is an arbitrary integer, this new topological number indexes only two topological phases, trivial and nontrivial.
Although the spin-orbit interaction in graphene was too weak to protect the edge states, soon after, the QSH effect based on the same mechanism has been predicted in HgTe/CdTe quantum wells\cite{Bernevig} and later confirmed experimentally.\cite{Koenig} Time reversal invariant (TRI) topological insulators were also predicted in 3D\cite{FuKaneMele,*MooreBalents,*Roy,*FuKane} and  soon reported experimentally. \cite{Hsieh08}
A complete classification of topological phases of one-particle fermionic insulating Hamiltonians in any dimension has been achieved.\cite{Kitaev,*Schnyder,*StoneSym,*Abramovici} For comprehensive reviews of this developing field, we point the interested reader to recent reviews.\cite{HasanRMP, *QiRMP}

In the present paper, we focus on the building blocks of the 2D 4-band QSH insulator, namely the 2-band Chern insulator. We first want to point out that not all QSH insulators can be  decomposed into two copies of Chern insulators. Such decomposition is only possible when the conservation of the third spin component $s_3$ is ensured. In such a case, the $\mathbb{Z}_2$ invariant is a \emph{spin Chern number} defined by the difference of the topological indices of the two Chern insulators formed by each spin species.\cite{Sheng} Based on geometrical considerations, 
we first provide a simple analytical method and formula to compute 
the topological index of a 2-band Chern insulator. 
Then we apply this approach to engineer general Hamiltonians for 2-band Chern insulators. 
We show in particular how to create models with an arbitrary large Chern number, $\text{Ch}$, with only two bands.
In comparison to the original Haldane model,\cite{Haldane} a Chern number is considered as large when $|\text{Ch}|\geq 2$.\cite{HCChen,*Jiang} 

The paper is organized as follows. In  Sec.~\ref{sec:chern}, we describe a general procedure to analytically calculate Chern numbers for 2-band insulators
based on the notion of degree of a map and test it on some well-known Chern insulators.
In Sec.~\ref{sec:model} we demonstrate how high Chern insulators can be engineered. In particular a new topological insulator with five available phases is constructed. 
Sec.~\ref{sec:buledge} is devoted to edge states, appearing in the presence of boundaries. Transitions between different topological phases are reflected 
in the structure of the edge states on a finite cylindrical geometry. This issue is explored both numerically and analytically. Finally, in Sec.~\ref{sec:Z2}, we discuss the realization of $\mathbb{Z}_2$-insulator by ``doubling'' the proposed model. We display the type of one-particle TRI perturbation that gaps the edges of the system composed of models in Chern number $\pm 2$ phases.

\section{Efficient determination of the topological index for 2-band Chern insulators}
\subsection{Chern number as a finite sum}
\label{sec:chern}
A gapped 2D system with TRS breaking can be characterized by the Chern number of the occupied bands given by\cite{Avron}
\begin{equation}\label{proj}
\text{Ch}=\frac{i}{2\pi}\int_{\rm BZ}\tr(dP\wedge PdP),
\end{equation}
where $P$ is the projector on the occupied bands. For a translationally invariant system, the Chern number is a topological index characterizing the $U(1)$ vector bundle over the first Brillouin zone (BZ). 
A non-zero Chern number can be understood as an obstruction for a global gauge choice for the wave function on the BZ. Physically this phenomenon is directly related to the quantization of the Hall conductivity $\sigma_H=\text{Ch}\times e^2/h$, where ${\rm Ch}$ corresponds to the number of edge states.\cite{Thouless,Hatsugai}

Any 2-band Hamiltonian can be expanded in term of pseudo-spin Pauli matrices $\sigma_i$ such that
\begin{equation}
H(\mb k)=\sum_{i=0}^3 h(\mb k)_i \sigma_i, 
\end{equation}
where $\sigma_0$ is the identity matrix. The term $h_0\sigma_0$ just shifts identically the energy bands and does not modify the topology of the Hamiltonian. We should therefore neglect it and write the Hamiltonian as $H(\mb k)=\mb h(\mb k)\cdot\bm\sigma$. This form reminds that of a Zeeman Hamiltonian for a spin $1/2$ particle in a magnetic field, except that the ``field'' $\mb h(\mb k)$ is defined in momentum space.

In the case the 2-band model the projector on the occupied band reads $P=(\sigma_0-\hat{\mb h}\cdot\bm\sigma)/2$, where $\hat{\mb h}=\mb h/|\mb h|$. Then from Eq.~(\ref{proj}), it follows that Chern number for the occupied band reads
\begin{equation}\label{Chern}
\text{Ch}=\frac{1}{4\pi}\int_{\rm BZ}d^2\mb k\,\hat{\mb h}
\cdot(\pd_{k_x}\hat{\mb h}\times\pd_{k_y}\hat{\mb h}).
\end{equation}

The Chern number is a winding number that counts how many times does the surface traced by $\mb h$
wrap around the origin $(0,0,0).$\cite{Stone}
The only practical difficulty in determining the Chern number lies in performing the integration in Eq.~(\ref{Chern}). We propose instead to calculate ${\rm Ch}$  using a discrete summation by determining directly the Brouwer degree of the map $\hat{\mb h}$.\cite{Milnor,Dubrovin}

Notice that because the vector $\mb h$ describes an insulator with a gap at zero energy, $\mb h(\mb k)$ is non-zero for all $\mb k$ in BZ. Then $\mb h(k_x,k_y)=(h_1,h_2,h_3)$ will trace a closed parametric surface, such that $\mb h: T^2\to\mathbb{R}^3\backslash\{(0,0,0)\}$ is a mapping from the Brillouin zone torus $T^2$ to a two dimensional surface $\mathcal M$ traced by the vector $\mb h(T^2)$. Then $\hat{\mb h}$ can be thought of as a composition $\hat{\mb h}=\pi\circ\mb h$, where $\pi:\mathbb{R}^3\backslash\{(0,0,0)\}\to S^2$ is the central projection to the unit sphere.

The condition to calculate the degree of $\hat{\mb h}$ are met: $T^2$ and $S^2$ are orientable manifolds without boundary and have the same dimension, $T^2$ is compact and $S^2$ is connected. In the general case, for a point $\mb k$ in $T^2$ one defines the derivative map between tangent vector spaces $d\hat{\mb h}(\mb k):T_{\mb k}T^2\to T_{\mb y}S^2$. Let $\sgn\,d\hat{\mb h}(\mb k)$ stand for the sign of the corresponding Jacobian at $\mb k$.
Then Chern number is equal to the Brouwer degree 
\begin{equation}\label{jac}
\text{Ch}=\sum_{\mb k\in \hat{\mb h}^{-1}(z)}
\sgn\,d\hat{\mb h}(\mb k),
\end{equation}
where $z$ is a regular point on the unit sphere.

Chern number can also be computed in terms of $\mb h$ in the following way. Consider the set $Y=\mathcal M\cap\pi^{-1}(z)$. Then one has
\begin{equation}\label{chrn}
\text{Ch}=\sum_{\mb y\in Y}\sum_{\mb k\in\mb h^{-1}(\mb y)}
\sgn[(\pd_{k_x}\mb h\times\pd_{k_y}\mb h)\cdot\mb n],
\end{equation} 
where the $\mb n$ is the unit vector towards $z$. This expression is just the generalization of the calculation of the winding number for a closed curve in 2D wrapping around a point $p$.\cite{Dubrovin,WangDeg}

The formula in Eq.~(\ref{chrn}) can be further simplified by an appropriate choice of the point $z$. The central projection $\pi$ maps any intersection point between $\mathcal M$ and a ray originating at $(0,0,0)$ to $z$.  If this ray does not cross the surface $\mathcal M$ traced by $\mb h$, it follows immediately that the surface does not wrap around the origin and the Chern number is zero. For a point $z$ on $S^2$ one can immediately obtain a set of points on $\mathcal M$ that project to $z$ through $\pi$. Since the expression~(\ref{jac}) does not depend on $z$, the choice of the latter can be guided by convenience. For instance, one can consider $z$ lying at a coordinate axis. Let us choose for example the $\sigma_3$-axis which intersects $\mathcal M$ in a set of points. This is equivalent to say that the components on the other axes are zero. Said differently,
the intersection of $\mathcal M$ with the  $\sigma_3$-axis are images of the band touchings originating from the simplified Hamiltonian $\sigma_1 h_1+\sigma_2 h_2$. Consequently, instead of integrating over the entire BZ, one only needs to consider the band touchings of the simplified Hamiltonian. Note that if $h_3$ is also zero at these points, then the system is not in a gapped phase.

Now it only remains to account for the orientation of the surface at the intersection points with $\sigma_3$-axis. This can be done by studying the projection of the surface normal vector on the $\sigma_3$-axis $(\pd_{k_x}\mb h\times\pd_{k_y}\mb h)_3$. When $h_3>0$, assume that the orientation is $(+1)$ when the sign of the projection is positive and $(-1)$ when the sign is negative; the converse is true when $h_3<0$. Then finding the Chern number amounts to a computation of a finite sum. Since we consider the entire $\sigma_3$-axis instead of a ray, the sum yields Chern number multiplied by two.

Note that although the choice of $\sigma_3$-axis is arbitrary, one must assure that the Jacobians in Eq.~(\ref{jac}) are all non-zero.
This is equivalent to say that the band touchings are Dirac points for the simplified (two Pauli matrices) Hamiltonian. In contrast, for a gapless point with a quadratic dispersion, $(\pd_{k_x}\mb h\times\pd_{k_y}\mb h)$ is zero as $\pd_{k_x}\mb h$ is collinear to $\pd_{k_y}\mb h$.\cite{Asano} If this is the case a different point $z$ should be used.

We can generalize and summarize the above argument in a formula for Chern numbers describing 2-band systems
\begin{equation}\label{ChForm}
{\rm Ch}=\frac{1}{2}\sum_{\mb k\in\mb D_i}
\sgn\big(\pd_{k_x}\hat{\mb h}\times\pd_{k_y}\hat{\mb h}\big)_i\sgn(h_i).
\end{equation}
where $i$ is the arbitrary axis chosen in (pseudo-)spin space.
Therefore the integral over momenta $\mb k$ in the BZ becomes a sum over $\mb k$ in the set of Dirac points $\mb D_i$ for Hamiltonians $H[h_i=0]$ (where $H$ is the original Hamiltonian $\mb h\cdot\bm\sigma$). Note that division by two is required because we considered the entire axis instead of a ray originating in $(0,0,0)$.

Eq.~(\ref{ChForm}) allows for an efficient determination of the Chern index and will guide us in constructing Hamiltonians with arbitrary Chern numbers.
Eq.~(\ref{ChForm}) also tells us that one can start with a gapless system \textit{containing only two $\sigma$ matrices} with bands that have Dirac points (a graphene like Hamiltonian for example). Afterwards one needs to compute the sign of the Jacobian $J_i$ defined by
\begin{equation}\label{jaco}
J_i=(\pd_{k_x}\mb h\times\pd_{k_y}\mb h)_i,
\end{equation}
at the Dirac points. It contains the information about the Berry phase gained by a particle moving around the Dirac points. Note again that if the gapless points do not have the characteristic linear dispersion, $J_i$ is zero.\cite{Asano} The sign of $J_i$ gives the Dirac points ``chirality''. Since the surface $\mathcal M$ is closed, the number of Dirac points for the Hamiltonians with two $\sigma$ matrices is always even.

To engineer a Chern insulator it is necessary to gap the system by adding a ``mass term''. As we shall see, the sign of the mass term at each Dirac point can be chosen independently, while keeping the Hamiltonian local.
As argued above, a necessary condition to obtain topological phases is that such a mass term takes different signs at the Dirac points. For example, if the mass term does not change its sign, then the surface traced by $\mb h$ does not wrap around the origin and the system is trivially gapped (${\rm Ch}=0$). Note that this also implies the general statement that the sum over chiralities is zero. Finally it must be stressed that there is nothing special about this third term. Any of the other terms can play the role of mass for simplified Hamiltonians formed by the remaining ones.

Let us call a ``topological charge'' the product of chirality and mass sign. From Eq.~(\ref{ChForm}) we see that each Dirac point contributes with $\pm 1/2$ topological charge to the Chern number.

From the arguments above, it is clear that a prerequisite to obtain a  Chern number $n$ is starting with a Hamiltonian displaying at least $2n$ Dirac points. Then to reach the largest possible Chern number (here $\pm n$), one must add a mass term such that topological charges do not annihilate. For example if the mass sign is tuned to have always the same (opposite) value as the chirality at its respective Dirac points, one obtains maximal $n$ (minimal $-n$) Chern number. It is clear that by tuning such a mass term, all topological phases with a Chern number between $-n$ and $n$ are {\it a priori} possible. Note, however, that in general the real space realization of this term might require longer than the nearest neighbor hopping.

Graphene, for example contains, two Dirac points. To gap it, one can add a third term to the Hamiltonian that changes the sign between the Dirac points. As we shall see below this can yield the famous Haldane model.\cite{Haldane}

\subsection{Examples}

Before constructing a high Chern topological insulator, we illustrate the efficiency of Eq.~(\ref{ChForm}) to compute Chern numbers on a couple of popular models of topological insulators: Haldane model\cite{Haldane} and Bernevig-Hughes-Zhang\cite{Bernevig} (BHZ) ``spin up'' Hamiltonian for the HgTe/CdTe quantum wells.

\subsubsection{Haldane model}
Let us start by considering the paradigmatic Haldane model.\cite{Haldane} This model is built starting form the tight-binding Hamiltonian for graphene. Graphene is a hexagonal lattice built out of two inter-penetrating triangular sub-lattices A and B. It contains only nearest neighbor electron hopping with the hopping integral $t_1$. The Haldane model contains also a second nearest neighbor hopping $t_2$, such that when the hopping is performed clockwise in the unit cell the electron gains a phase $\phi$. However, the overall phase on the unit cell is zero; there is no net magnetic flux. Let the vectors $(\mb a_1, \mb a_2, \mb a_3)$ describe the displacements from B atoms to \emph{nn} A atoms and $\mb b_i=\frac{1}{2}\e_{ijk}(\mb a_j-\mb a_k)$ vectors relating \emph{nnn} sites.

The Bloch Hamiltonian reads
\begin{align}
H(\mb k)&=\sum_{i=1}^3\big\{2t_2\cos(\phi)\cos(\mb{k\cdot 
b_i})\sigma_0+t_1[\cos(\mb{k\cdot a_i})\sigma_1\notag\\
&\quad+\sin(\mb{k\cdot a_i})\sigma_2]+\big[\frac{M}{3}-2t_2\sin(\phi)\sin(\mb{k\cdot b_i})\big]\sigma_3\big\},\notag\\
\end{align}
where $\pm M$ is the on-site energy.
Let us study the topology of the surface traced by vector $\mb h$. Again, if one chooses to study the points where the surface intersects $\sigma_3$ axis, it follows immediately that those are exactly given by the Dirac points of graphene, i.e. by the Hamiltonian with $M\to 0$ and $t_2\to 0$. It is easier to reparametrize the Hamiltonian, such that $\mb{k\cdot b_3}=q_x$ and $\mb{k\cdot b_1}=q_y$. This describes a mapping to a square BZ, with Dirac points given by 
$\mb q^0_{\pm}=\pm (2\pi/3,2\pi/3)$. Because $\mb a_i=\frac{1}{6}\e_{ijk}(\mb b_k-\mb b_j)$, $\mb k\cdot(\mb{a_1,a_2,a_3})=-\frac{1}{3}
(-2q_x-q_y,q_x-q_y,q_x+2q_y)$.
Dirac points' chirality is given by the sign of the Jacobian in Eq.~(\ref{jaco}). 
We find
\begin{equation}
\sgn(J_3^{\pm})=\mp 1,
\end{equation}
where $J_3^{\pm}$ denotes the Jacobian calculated at the Dirac point 
$\mb q^0_{\pm}$.

In order to calculate the Chern number one must also consider the mass sign at the Dirac points, $m_{\pm}(\mb q^0_\pm)=M\mp 3\sqrt{3}t_2\sin(\phi)$. Therefore, following Eq.~(\ref{Chern}) one recovers Haldane result for the Chern number
\begin{equation}
\text{Ch}=\frac{1}{2}[\sgn(m_-)-\sgn(m_+)].
\end{equation} 

The transition from a topological insulator to a normal insulator is marked by a semi-metal state, where the gap closes at least at one Dirac point when $M=\pm 3\sqrt{3}t_2\sin(\phi)$. In the normal insulating phase Dirac points have identical associated mass sign, such that Ch goes to zero.

\subsubsection{BHZ model}
The Chern number calculation can as well be exemplified in the case of the recently discovered $\mathbb{Z}_2$ insulators such as the 2D HgTe/CdTe quantum wells.\cite{Bernevig} The low energy Bloch Hamiltonian is written in a basis of four states $|E1,m_J=1/2\ra$, $|H1,m_J=3/2\ra$, $|E1,m_J=-1/2\ra$, $|H1,m_J=-3/2\ra$ and it has the form
\begin{equation}
\tilde H=
\begin{pmatrix}
H(\mb k) & 0 \\
0 & H^*(-\mb k)
\end{pmatrix}
\end{equation}
The system is assembled out of two Chern insulators, $H(\mb k)$ and its time reversed copy. To illustrate the calculation of the Chern numbers using a discrete sum, it is enough to pick one Chern insulator:
\begin{eqnarray}
H(\mb k)&=&A\sin(k_x)\sigma_1+A\sin(k_y)\sigma_2\notag\\
&&+(M-2B(2-\cos(k_x)-\cos(k_y)))\sigma_3,
\end{eqnarray}
where $A,B,M$ are parameters of the Hamiltonian.
Let us  consider again the surface traced by $\mb h$ and choose  $\sigma_3$ as  a special axis.
The points where the $\sigma_3$-axis pierces the surface are given by the condition that $h_1$ and $h_2$ terms vanish. This gives four ``Dirac points'' $(k_x,k_y)\in\{(0,0),(0,\pi),(\pi,0),(\pi,\pi)\}$.

The chirality of each Dirac point is given by the sign of Jacobian $J_3$
\begin{equation}
\sgn(J_3)=\sgn[\cos(k_x)\cos(k_y)]
\end{equation}
evaluated at the Dirac points. The mass term $h_3$ has the following expression at the Dirac points, $h_3(0,0)=M$, $h_3(0,\pi)=h_3(\pi,0)=M-2B$ and $h_3(\pi,\pi)=M-4B$.

The Chern number can then be easily computed for different values of $M$ and $B$. The results for the case $B>0$ are summarized in Tab.~\ref{tab:chern}. 
\begin{center}
\begin{table}
\begin{tabular}{lccccc}
\hline\hline
Dirac points & (0,0) & ($\pi$,0) & (0,$\pi$) & ($\pi$,$\pi$) & Ch\\
\hline
mass & $M$ & $M-4B$ & $M-4B$ & $M-8B$\\
\hline
chirality & + & $-$ & $-$ & +\\
\hline
$M<0$ & $-$ & $+$ & $+$ & $-$ & 0\\
$M\in(0,4B)$ & $+$ & $+$ & $+$ & $-$ & +\\
$M\in(4B,8B)$ & $+$ & $-$ & $-$ & $-$ & $-$\\
$M>8B$ & $+$ & $-$ & $-$ & $+$ & 0\\
\hline\hline
\end{tabular}
\caption{Values of the Chern number $\rm Ch$ according to the different values taken by $M$ and $B>0$}
\label{tab:chern}
\end{table}
\end{center} 
It follows from Tab.~\ref{tab:chern} that as $M$ varies between 0 and $8B$, the Chern insulator $H(\mb k)$ exhibits two topologically nontrivial phases with ${\rm Ch}=\pm 1$. When $M$ is outside $(0,8B)$ region there is only a trivial insulator phase. Note that with respect to $H(\mb k)$, its time reversed copy will always have the opposite Chern number.

\section{\texorpdfstring{Higher Chern number models. Construction of a Chern insulator with phases $\{0,\pm 1,\pm 2\}$}{Higher Chern number models}}
\label{sec:model}
In the following section we present the construction of a 2D model with a higher Chern number. The strategy will be to work with an abstract Hamiltonian in  $k$-space. We delay its implementation on a particular lattice at the end of this section. New topological phases are made possible by tuning the parameters of the mass term and controlling the addition and subtraction of ``topological charges''.

\subsection{Topological phases in $k$-space Hamiltonian}
The most important common point between Haldane model and ``half''-BHZ Chern insulators is the existence of only three topological phases described by Chern numbers in $\{0,\pm 1\}$ ($0$ characterizing the trivial phase). 
Using Eq.~(\ref{ChForm}) it is simple to construct an artificial model which has larger Chern number. 
We assume that we are in momentum space with a general 2D Hamiltonian of the type
\begin{equation}
h_1(k_x,k_y)\sigma_1+h_2(k_x,k_y)\sigma_2.
\end{equation}
$h_1$ and $h_2$ are continuous, $2\pi$ periodic functions of $k_x$ and $k_y$.

The bands meet when both functions $h_{1}$ and $h_2$ are zero. 
This defines a system of two equations with two variables $k_{x},k_{y}$, whose solutions define points in $k$-space. 
As explained above, in order to obtain a higher Chern number (here $\pm 2$) it is necessary to have at least four Dirac points.

Let us start by considering the basic template on which the topological insulator is built. 
Among the simplest models, one can consider the following Hamiltonian:
\begin{equation}
H_0(k_x,k_y)=2t_1[\cos(k_x)\sigma_1+\cos(k_y)\sigma_2].
\end{equation}
Here, we assume that the Pauli matrices $\sigma$ correspond to some   pseudo-spin or orbital degree of freedom. Note that already the system breaks TRS, $TH_0(\mb k)T^{-1}\neq H_0(-\mb k)$.

The energy dispersion reads
\begin{equation}
E=\pm 2t_1\sqrt{\cos^2(k_x)+\cos^2(k_y)},
\end{equation}
such that there are four Dirac points $\mb k_0=(\pm\pi/2,\pm\pi/2)$.

The Dirac point chirality is given by the sign of the Jacobian $J_3$
\begin{equation}
\sgn(J_3)=\sgn[\sin(k_x)\sin(k_y)].
\end{equation}
This determines immediately the chirality $\chi$ of the four Dirac points as summarized in Tab. \ref{tab:chirality}.
\begin{center}
\begin{table}
\begin{tabular}{ccccc}
\hline\hline
Dirac points & $(\frac{\pi}{2},\frac{\pi}{2})$ & $(\frac{\pi}{2},-\frac{\pi}{2})$ & $(-\frac{\pi}{2},\frac{\pi}{2})$ & $(-\frac{\pi}{2},-\frac{\pi}{2})$ \\[2pt]
\hline
$\chi$ & + & $-$ & $-$ & + \\
\hline\hline
\end{tabular}
\caption{Chirality $\chi$ of the four different Dirac points}
\label{tab:chirality}
\end{table}
\end{center}

Notice that Dirac points at $\mb k$ and $-\mb k$ have the same chirality. They will be referred in the following as a pair of Dirac points.

To obtain a Chern insulator one needs to add a mass term. As we can see from Eq. (\ref{ChForm}), the Chern index depends on both the chiralities of the Dirac points and the sign of the mass term in their vicinity. Let us add a mass term of the form $h_3(k_x,k_y)\sigma_3$. Since $h_3$ is a periodic function on the BZ, in the general case its zeros form a set of closed lines on the two-dimensional torus. The first condition in order to gap the initial system is that these lines must not pass through the Dirac points. Thus the lines of zeros delimit regions, $\mathcal R_{1}$ and $\mathcal R_{2}$, where the mass term has the same sign. For a proper choice of the gap term it is somewhat simpler to see the problem from a geometrical point of view. In order to maximize $|\text{Ch}|$, one needs that the lines of zeros separate the \emph{pairs} of Dirac points such that a pair of points of a given chirality are contained in a region of positive mass, while pair of points of opposite chirality are contained in a negative mass region. In short, each \emph{pair} of Dirac points are placed in regions where the mass term has different values. On the contrary, if a pair of Dirac points is ``broken'' (such that one is in region $\mathcal R_1$ and the other in region $\mathcal R_2$), then the topological charges will cancel out as  can be directly inferred from Eq.~(\ref{ChForm}). 

\begin{figure}[t]
\centering
\subfloat{\label{fig:FBZ1}\includegraphics[width=0.4\textwidth]{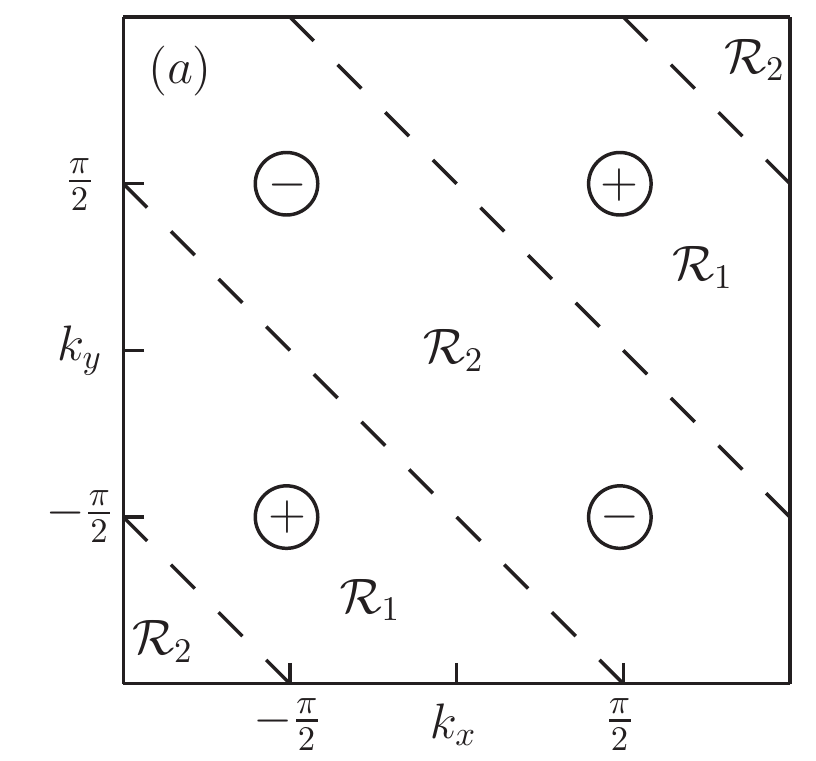}}\\
\subfloat{\label{fig:FBZSin}\includegraphics[width=0.4\textwidth]{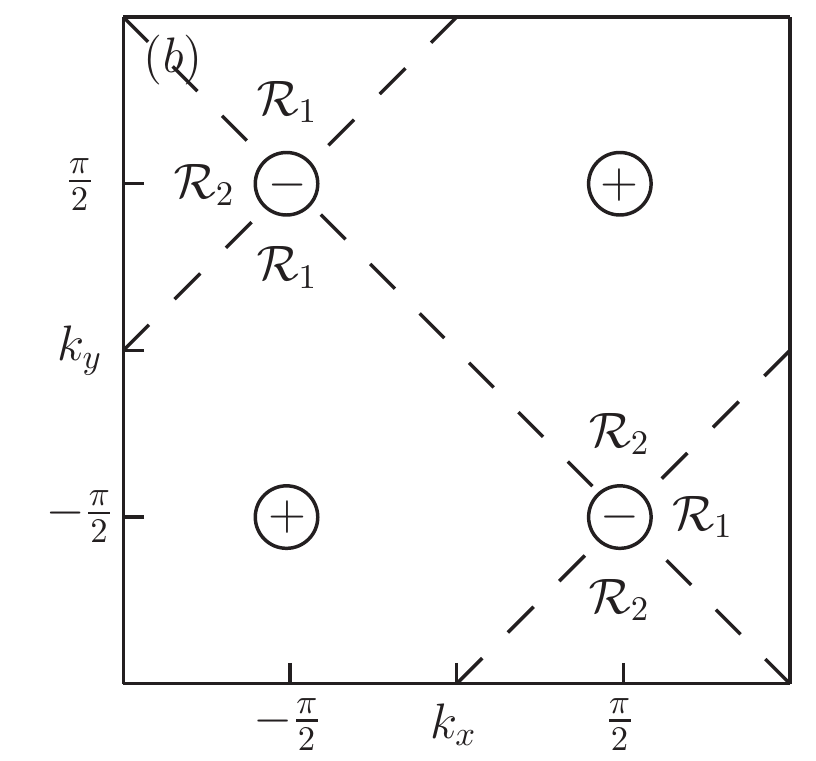}}
\caption{(a) Domains associated to the mass term  $\cos(k_x+k_y)$. Phases with Chern number $\pm 2$ are realized.
(b) Domains associated to the mass term $\sin(k_x)+\sin(k_y)$. Phases with Chern number $\pm 1$ are possible after the remaining Dirac points are identically gapped.
In both figures, 
dashed lines denote the locus of points of zero mass term, while $\mathcal R_1$ and $\mathcal R_2$ denote regions with opposite values for the mass term. The solid circles represent Dirac points of the initial gapless system; the corresponding chiralities are marked by the signs within.}
\end{figure}

Let us first realize the topological phases with a Chern number $\pm 2$ where each pair of Dirac points is in different regions $\mathcal R$. The periodic function $\cos(k_x+k_y)$ accomplishes such demand with lines of zeros  given by $k_y=-k_x+\frac{2n+1}{2}\pi$. The regions $\mathcal R_1$ and $\mathcal R_2$ for this term are represented in Fig.~(\ref{fig:FBZ1}). The mass term is negative in region $\mathcal R_1$ and positive in region $\mathcal R_2$. The lowest Chern number $\text{Ch}=-2$ is obtained for this model. Note that the above choice of $h_3$ is not completely arbitrary, since the term $t_2\cos(k_x+k_y)\sigma_3$ corresponds to a short range hopping with the amplitude $t_2$. Hence, the modified Hamiltonian $H_1$ becomes
\begin{equation}\label{eq:h1}
H_1(\mb k)=2t_1\cos(k_x)\sigma_1+2t_1\cos(k_y)\sigma_2+2t_2\cos(k_x+k_y)\sigma_3.
\end{equation}
For this Hamiltonian, one readily computes $\text{Ch}=2\sgn(-t_2)$.
Therefore the sign of $t_2$ determines whether a maximal or minimal Chern number is reached. Note that the Hamiltonian in Eq.~(\ref{eq:h1}) lacks trivial insulating phases or Chern phases with $\text{Ch}=\pm 1$. 

To produce a trivial phase with $\text{Ch}=0$, it suffices to add a large, ``staggered potential'' $m\sigma_3$ such that the entire mass term for all Dirac points has the same sign.
Note that since Pauli matrices can act on any degree of freedom, the term ``staggered'' could for instance refer to the fact that two sites in a cell or two orbitals on a site have an associated $\pm m$ constant energy. 

The Hamiltonian then changes from $H_1$ to $H_2=H_1+m\sigma_3$ and consequently the Chern number becomes
\begin{equation}
{\rm Ch}=\sgn(-m-2t_2)+\sgn(m-2t_2).
\end{equation}
Therefore, when the mass is large enough, $|m|>2|t_2|$, the system enters a trivial phase. We have therefore a transition from
a $\text{Ch}=\pm 2$ phase to a $\text{Ch}=0$ phase.

Along the same reasoning, we can obtain phases with  $\text{Ch}=\pm 1$. Remember that Dirac points at $\mb k$ and $-\mb k$ have the same chirality. Then to obtain a $\text{Ch}=\pm 1$ it is necessary that the mass term is allowed to take different signs for exactly one pair of Dirac points. If one adds only even functions of $\mb k$ to the mass term, it would not be possible to ``break'' a pair of Dirac points. Consequently, an odd function 
is required. The simplest choice would be to add the term proportional to $\sin(k_x)+\sin(k_y)$. The mass for one pair of Dirac points is unchanged, while for the pair $(\mb k_0,-\mb k_0)$ with $\mb k_0=(\pi/2,\pi/2)$ the mass changes
(see Fig.~\ref{fig:FBZSin} for the model containing only $\sin(k_x)+\sin(k_y)$ in the mass term).
Note that as follows from Fig.~\ref{fig:FBZSin}, if the mass term contains solely $\sin(k_x)+\sin(k_y)$, the system is not an insulator because two Dirac points are not gapped. However, the presence of even functions in the mass term gaps identically the two Dirac points so that phases with Chern number $\pm 1$ become possible. Adding all the terms together gives the following complete Bloch Hamiltonian $H$
\begin{eqnarray}\label{BH}
H&=&2t_1\cos(k_x)\sigma_1+2t_1\cos(k_y)\sigma_2\nonumber\\
&&+[m+2t_2\cos(k_x+k_y)+2t_3(\sin(k_x)+\sin(k_y))]\sigma_3.\notag\\
\end{eqnarray}
There are four free parameters $(m,t_1,t_2,t_3)$ in our model, and five insulating topological phases $\{0,\pm 1,\pm 2\}$. Let us assume that all parameters are real. All the phases can be reached by varying $(m,t_2,t_3)$ while keeping $t_1$ fixed. The Chern number for the final model reads
\begin{eqnarray}\label{ChF}
\text{Ch}&=&\sgn(-m-2t_2)+\frac{1}{2}[\sgn(m-2t_2+4t_3)\notag\\
&&+\sgn(m-2t_2-4t_3)].
\end{eqnarray}

\begin{figure}[t]
\centering
\includegraphics[width=0.35\textwidth]{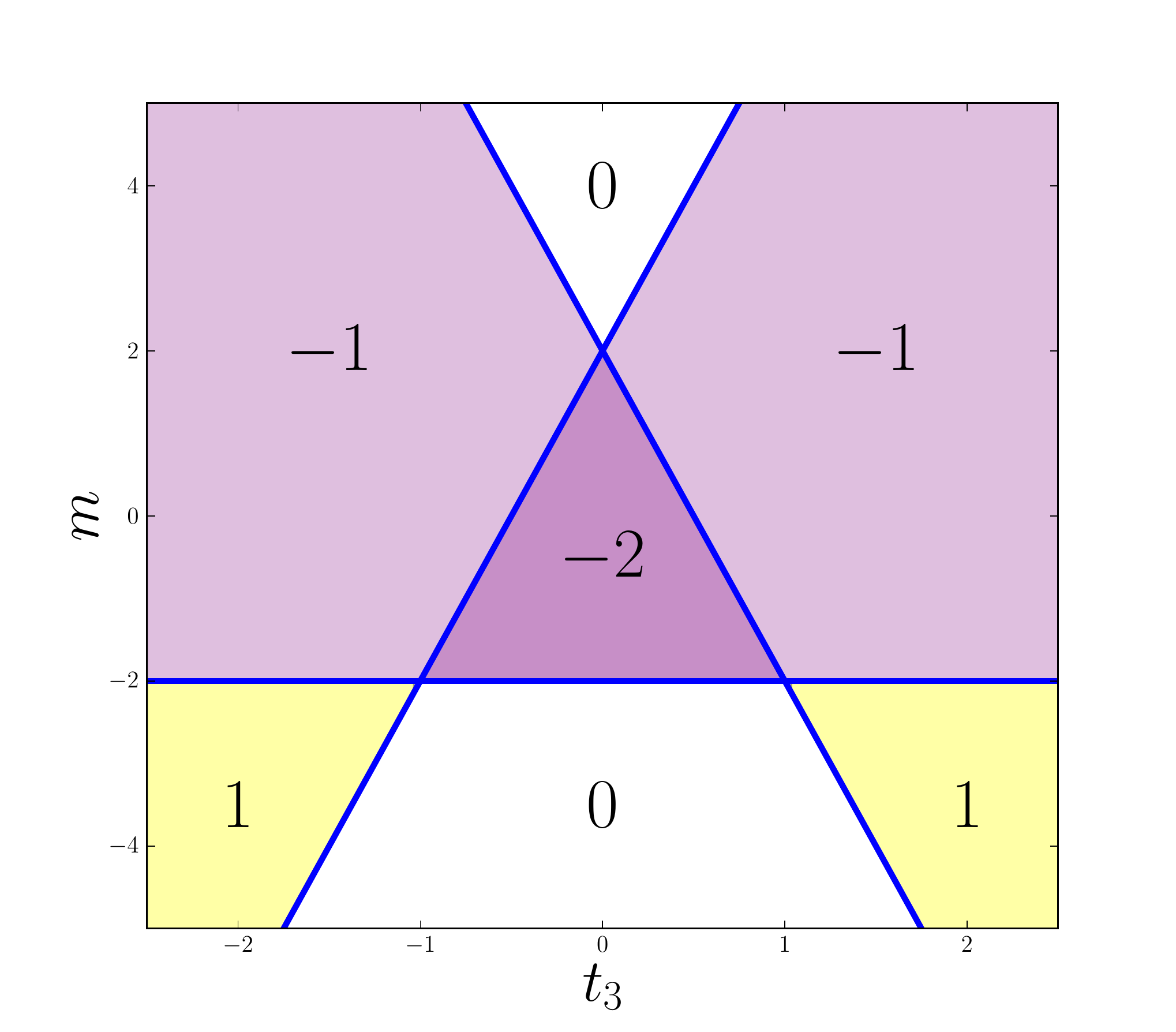}
\caption{Phase diagram of the system for $t_2=1$. Each region is denoted by the corresponding Chern number and represents an insulating topological phase. The boundaries of the regions represent topological transitions where the system becomes gapless. (color online)}
\label{fig:PhDiag}
\end{figure}

Such a formula gives immediately the phase diagram associated with the system described by the Hamiltonian in Eq.~(\ref{BH}). Notice that the parameter $t_1$ does not enter in the description of the phases, but manifestly needs to be finite to have non-vanishing $\sigma_1$ and $\sigma_2$ components in the model. The formula ($\ref{ChF}$) can be illustrated by the phase diagram on the Fig.~(\ref{fig:PhDiag}). This diagram contains only four phase of the model; the phase with ${\rm Ch}=2$ needs $t_2<0$.

\subsection{Direct space realization}
For the moment the system was assumed translationally invariant in 2D and was described abstractly in parameter space. However to investigate the edge states it is necessary to propose a lattice implementation for the model.

The system could be realized on a triangular lattice with two orbitals on each site. The parameters are interpreted as $k_{x,y}=\mb k\cdot\mb a_{1,2}$ with $\mb a_1$ and $\mb a_2$ the two Bravais vectors. The Pauli matrices are then  operating in the orbital space.

The Hamiltonian (\ref{BH}) can be rewritten as
\begin{eqnarray}\label{real}
H&=&\sum_{ij}[c^\dag_{ij}\frac{m}{2}\sigma_3c_{ij}+c^\dag_{i+1j}(t_1\sigma_1+i t_3\sigma_3)c_{ij}\notag\\
&&+c^\dag_{ij+1}(t_1\sigma_2+i t_3\sigma_3)c_{ij}
+c^\dag_{i+1j+1}t_2\sigma_3 c_{ij}+\text{h.c.}],\notag\\
\end{eqnarray}
where 
\begin{equation}
c_\mb k=\frac{1}{\sqrt{N}}\sum_{\mb r_{ij}} c_{ij}e^{-i\mb k\cdot\mb r_{ij}}
\end{equation}

\begin{figure}[t]
\centering
\includegraphics[width=0.3\textwidth]{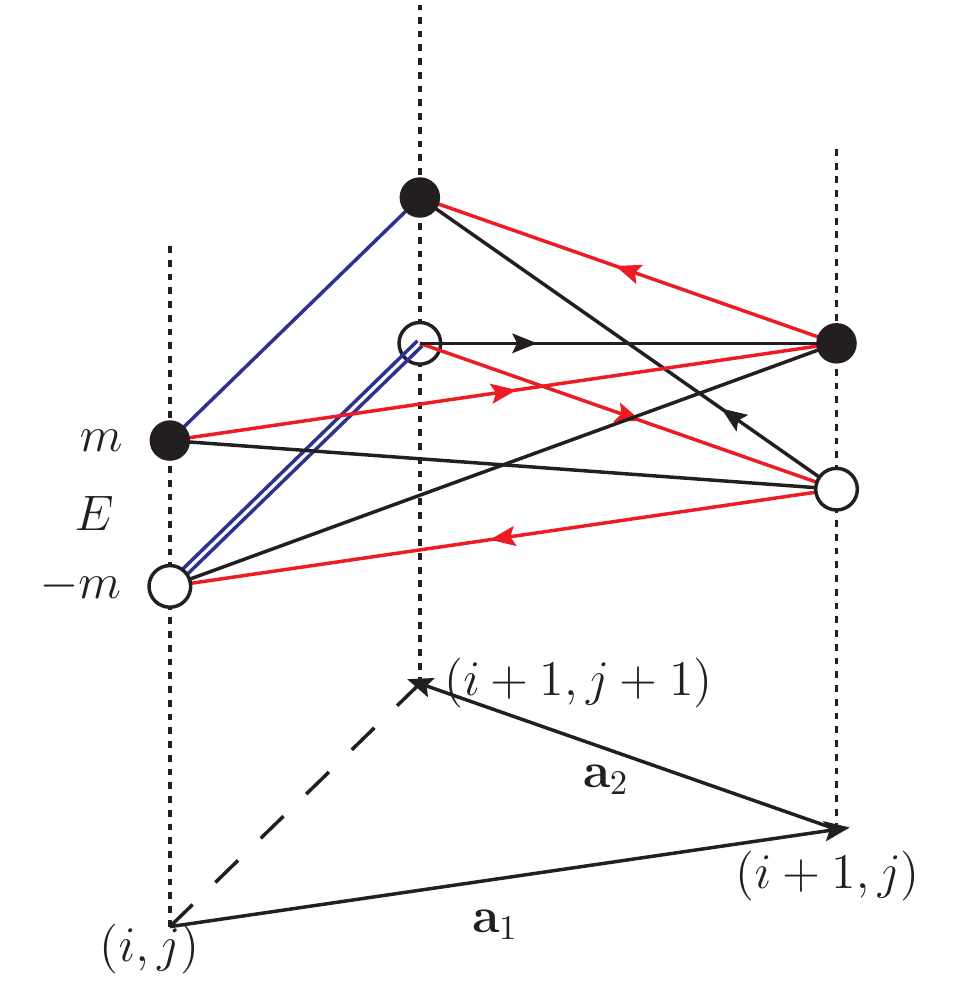}
\caption{Direct space realization of the model (\ref{BH}), see Eq.~(\ref{real}). $(i,j)$ denotes lattice sites.
 $\mb a_{1/2}$ are Bravais lattice vectors. $\bullet$ $(\circ)$ represents orbitals with energy $m$ ($-m$). The vertical axis represents the on-site energy difference between the two inequivalent orbitals. Black lines represent $t_1$ hoppings, blue lines, $t_2$ hoppings, and red lines, $t_3$ hoppings. An arrow on a link indicates that an electron hopping in the corresponding direction gains a $\pi/2$ phase. Similarly a double line indicates a $\pi$ phase gain. (color online)}
\label{fig:Tri}
\end{figure}

The position is defined by $\mb r_{ij}=i\mb a_1+j\mb a_2$ with $\mb a_{1/2}$ as Bravais vectors making an angle $2\pi/3$ between them (see Fig~(\ref{fig:Tri})). We set the lattice constant $a=1$. 
On each site there are two different orbitals with energy $\pm m$. The Hamiltonian (\ref{real}) describes through $t$-terms the overlap between (in)equivalent orbitals. The model is represented in Fig.~(\ref{fig:Tri}). Note that there is no net flux perpendicular to the two dimensional plane. But TRS is broken, because for certain closed paths the electron can still gain a non-zero phase.

\section{Bulk-edge correspondence}
\label{sec:buledge}
Until now, only bulk properties of the model were explored. In the present section, we investigate how these properties manifest themselves through the edge state structure. 
It was proved for general 2D Chern insulators that the Chern number is reflected in the number of gapless edge states one encounters on a finite geometry.\cite{Qi}

We have therefore implemented the aforementioned system on a cylinder and investigate subsequently the edge state wave functions and edge state dispersion at its two ends. In the first subsection, mainly numerical results are presented. It illustrates the fact that the Chern number reflects itself in the number of gapless states at the interface between phases with $\text{Ch}=-1$ and $\text{Ch}=-2$. In the second subsection, analytical methods are employed to obtain the edge state structure. Here the main goal is to obtain the edge wave function and dispersion for any parameter values. In particular, the complete edge state determination for a choice of parameters in high Chern $-2$ phase is given.

\begin{figure*}[t]
\centering
\subfloat{\label{fig:Transa}\includegraphics[width=0.25\textwidth]{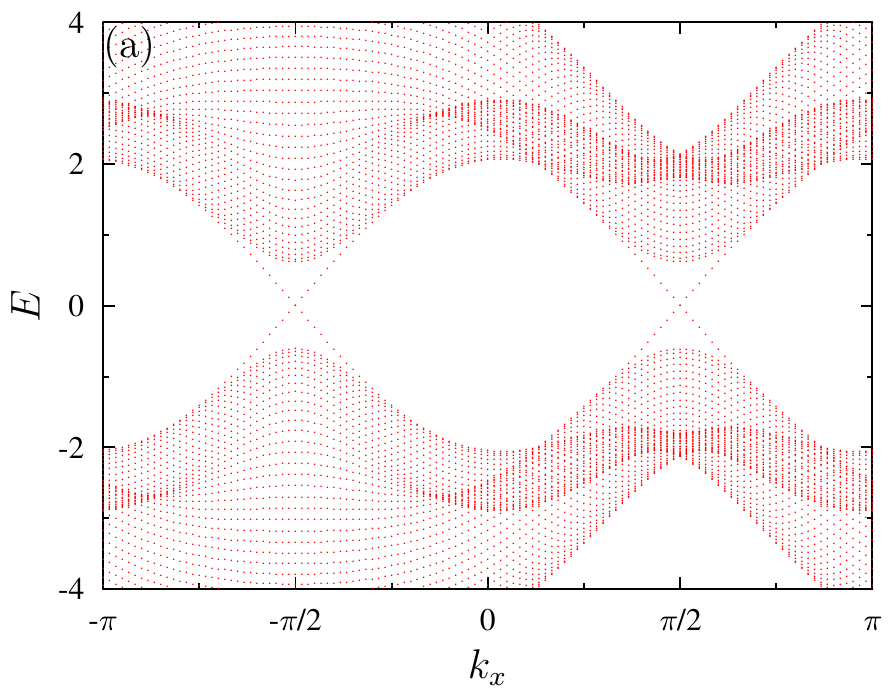}}
\subfloat{\label{fig:Transb}\includegraphics[width=0.25\textwidth]{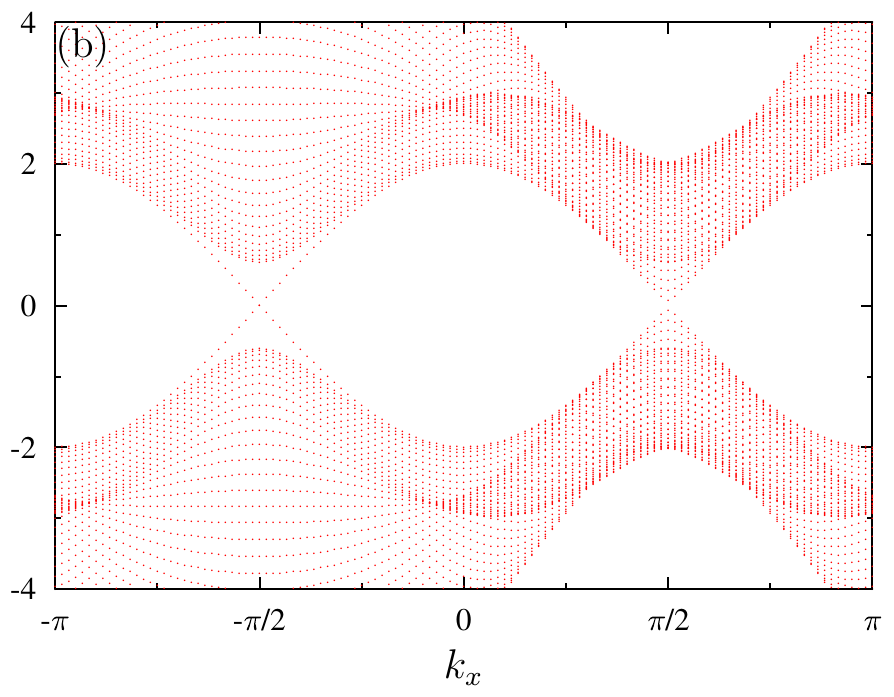}}
\subfloat{\label{fig:Transc}\includegraphics[width=0.25\textwidth]{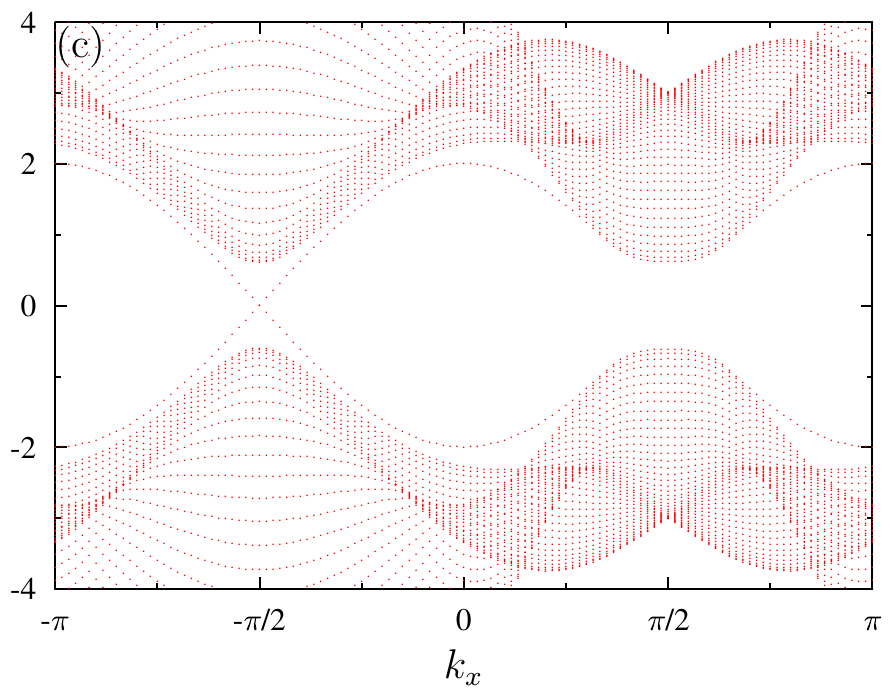}}
\subfloat{\label{fig:Transd}\includegraphics[width=0.22\textwidth]{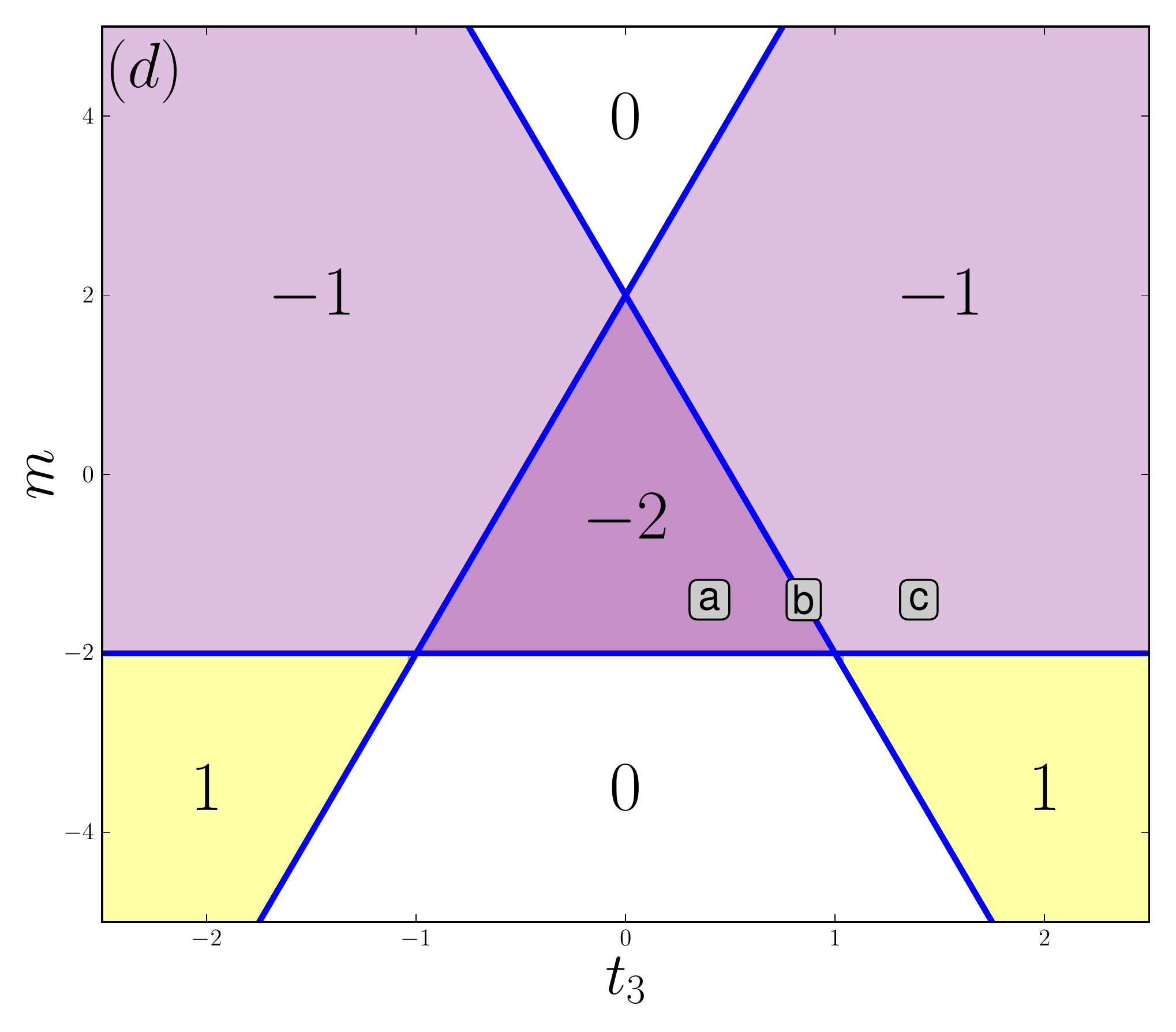}}
\caption{Energy dispersions for (a) $t_3=0.4$ ($\text{Ch}=-2$), (b) $t_3=0.85$ (topological phase transition at closing bulk gap), (c) $t_3=1.6$ ($\text{Ch}=-1$). The other parameters are $t_1=1$, $t_2=1$, $m=-1.4$. The  simulation is done when the system is on cylinder with height of 40 sites and a circumference of 180 sites. The number of edge states is $2\times|\text{Ch}|$ because there are two edges. (d) Representation of chosen points on the phase diagram. (color online)
}
\label{fig:Trans}
\end{figure*}
\subsection{Numerical experiments}
The finite geometry chosen for the numerical study is a cylinder. It is constructed out of a patch of the lattice having the shape of a parallelogram Bravais vectors $\mb a_1$ and $\mb a_2$ as edges. Subsequently the edges parallel to $\mb a_2$ are glued together to obtain the the final cylindrical shape. Because translational invariance is maintained in the direction parallel to $\mb a_1$, $k_x$ remains a good quantum number. Therefore, one can fix $k_x$ and consider the resulting one-dimensional problem. Let us write  the one-particle solutions of the corresponding stationary Schr\"odinger equation as 
\begin{equation}
|\psi(k_x,j)\rangle=\sum_{k_x}\psi_j(k_x)c^\dag_{k_x,j}|0\rangle,
\end{equation}
where $j$ denotes the layers of sites in $\mb a_2$ direction, and $\psi_j$ is a spinor due to the fact that there are two orbitals in the problem.

Then for a given quasi-momentum $k_x$, the Schr\"odinger equation reads
\begin{equation}\label{edgeEq}
\Gamma_1\psi_j+\Gamma_2\psi_{j+1}+\Gamma_2^\dag\psi_{j-1}=E\psi_j,
\end{equation}
where
\begin{eqnarray}
\Gamma_1&=&m\sigma_3+2t_1\cos(k_x)\sigma_1+2t_3\sin(k_x)\sigma_3\notag,\\
\Gamma_2&=&t_1\sigma_2+(t_2e^{ik_x}-it_3)\sigma_3.
\end{eqnarray}

The cylinder has the edges at $j=1$ and $j=L_y$.
Hard wall boundary conditions are imposed, such that the amplitudes $\psi_0$, $\psi_{L_y+1}$ must vanish. The dispersion $E$  as a function of $k_x$ is obtained by numerically solving Eq.~(\ref{edgeEq}) for the given boundary conditions and for different choices of the parameters (see Fig.~(\ref{fig:Trans})). In our numerical experiments the cylinder circumference is $L_x=180$ sites and with height $L_y=40$ sites. All the energies are measured in units of $t_1$.

To illustrate the bulk-edge correspondence, we sample several regions of the phase diagram on Fig.~($\ref{fig:PhDiag}$). In non-trivial topological regions, edge states appear around the ends of the cylinder. From the phase diagram we can predict the number of edge states in our model. For example, three sets of parameter values along the constant $m=-1.4$ line are taken such that the transition between the $-1$ and $-2$ phases is explored (see Fig.~(\ref{fig:Transd})). 

While the bulk remains insulating there are states crossing the gap. These are the edge states and their total number is $2\times|\text{Ch}|$ since the cylinder has two edges. Note that the edge states at zero energy, cross the gap at $k_x=\pm\pi/2$. At any topological transition the bulk closes at least in one of the special points $k_x=\pm\pi/2$. A transition changing the Chern number by two requires that the gap closes at both points, while for a change of one, only one Dirac cone forms.

\subsection{Analytical solution}
A greater insight into the model is gained by solving Eq.~(\ref{edgeEq}) analytically. It allows one to clearly identify the gapless states as edge states and also to determine their penetration length. The edge state dispersion law can be found either by a direct solution of Schr\"odinger equation or indirectly by the method described in Ref.~[\onlinecite{Mong}]. We explore both approaches in the general setting of our model and exemplify the results for a particular choice of parameters corresponding to the phase ${\rm Ch}=-2$.

As it was elegantly proved in Ref.~[\onlinecite{Mong}] the  condition of existence and edge state dispersion can, under certain provisions, be found from a simple analysis of bulk Bloch Hamiltonians. The method developed there applies when an infinite ribbon or a cylinder is cut out of the infinite 2D system. The direction of the cut must follow a Bravais lattice vector. In this case the momentum parallel to the cut $k_\parallel$ is conserved and the system splits into a set of 1D Hamiltonians describing the motion of the electron between the layers of sites parallel to the cut. The final prerequisite to apply the method is that there are only nearest neighbor layer hopping terms.  Eq. (\ref{edgeEq}) shows that it is indeed the case in the present model with $k_\parallel=k_x$.

According to Ref.~[\onlinecite{Mong}], the key information about the edge states can be revealed by studying the curves traced by $\mb h$ as a function of $k_\perp$ with fixed $k_\parallel$. In the case when only the nearest neighbor interlayer hopping is allowed these curves are planar (actually they are ellipses). Therefore, $\mb h$ can be decomposed in two parts, $\mb h_\perp$ perpendicular to the ellipse plane and $\mb h_\parallel$ the in-plane component. Each component yields some important piece of information about the edge states. Namely, the edge state with a given $k_\parallel$ exists if and only if the ellipse traced by $h_\parallel$ encloses the projection of the origin onto the plane of the ellipse. The energy of the state is equal to $\pm|\mb h_\perp|$.

In our case $k_\parallel=k_x$ and $k_\perp=k_y$. This yields 
\begin{eqnarray}\label{hpara}
\mb h_\parallel&=&
(0,2t_1\cos(k_y),2t_2\cos(k_x)\cos(k_y)\notag\\
&&+2(t_3-t_2\sin(k_x))\sin(k_y)+m+2t_3\sin(k_x)).\notag\\
\end{eqnarray} 
For a fixed $k_x$, the equation (\ref{hpara}) describes an ellipse parametrized by $k_y\in[0,2\pi)$. The condition that the ellipse encloses the origin reads
\begin{equation}\label{ECond}
|m+2t_3\sin(k_x)|<2|t_3-t_2\sin(k_x)|.
\end{equation}
This equation determines the range in $k_x$ where edge states exist.
The energy of the state is  $\pm 2t_1\cos(k_x)$.

Although the edge dispersion is determined, it must accommodate up to four edge states. The existence condition also gives for $|\text{Ch}|=2$ two intervals of allowed $k_x$ where edge states exist.

The edge states can also be explored directly by studying the special solutions of equation (\ref{edgeEq}). The strategy consists firstly in finding a zero energy solution. From the above analytical results or from numerical experiments it is apparent that if there are edge states, they will cross zero energy only at $k_x=\pm\pi/2$. After obtaining the solution at this particular $k_x$-points, one extends the solution for the entire range of $k_x$ which allows edge states. Note that although this method relies on the starting information about the zero energy crossing of edge states, it determines the existence condition and dispersion by itself.

Let us consider the solutions to Schr\"odinger equation (\ref{edgeEq}) of the form $\psi_j=\rho^j\phi$. The complex parameter $\rho$ captures the behavior of the wave function in the direction transverse to the edge $\rho=e^{ik_y}$ and $\phi$ carries the spinor structure of $\psi_j$. The ensuing equations, at $k_x=\pm\pi/2$ read
\begin{eqnarray}\label{eq:recurrence}
(0,0)^T&
=&\{[(t_1\sigma_2-i(t_3\mp t_2)\sigma_3)]\rho\notag\\
&&+[(t_1\sigma_2+i(t_3\mp t_2)\sigma_3)]\rho^{-1}
+(\pm 2t_3+m)\sigma_3\}\phi.\notag\\
\end{eqnarray}

By multiplying Eq.~(\ref{eq:recurrence}) with $\sigma_3$ on the left hand side, we obtain an equation which depends only on $\sigma_1$. Therefore $\phi$ is a linear combination of the eigenstates $|x\pm\ra$ of $\sigma_1$. The solution can then be extended beyond the special points $k_x=\pm\pi/2$ by continuity.

Eigenmodes $\rho$ are found from the general $k_x$-dependent equation
\begin{equation}
(\Gamma_1+\Gamma_2\rho+\Gamma_2^\dag\rho^{-1})|x\pm\ra=0.
\end{equation} 
Note that there are two solutions $\rho$ for a given eigenstate $|x\pm\ra$.
A certain symmetry between the solutions is immediately apparent. If $\rho_1$ and $\rho_2$ are solutions to the equation for the eigenstate $|x+\ra$, then $\rho^{*-1}_1$ and $\rho^{*-1}_2$ are solutions for the equation corresponding to $|x-\ra$ eigenstate. In particular, $\rho_1$ and $\rho_2$ associated to $|x+\rangle$, are determined from
\begin{eqnarray}
0&= &\,m+2t_3\sin(k_x)+t_2(e^{ik_x}\rho+e^{-ik_x}\rho^{-1})
-it_3(\rho-\rho^{-1})\notag\\
&&-it_1(\rho+\rho^{-1}).
\end{eqnarray}
Note that the edge state dispersion follows as a byproduct $E=\pm 2t_1\cos(k_x)$ depending on the eigenstate $|x\pm\ra$. 

For a given $k_\parallel$ the general solution to Eq. (\ref{edgeEq}) thus reads
\begin{equation}
\psi_j=(c_1\rho^{j}_1+c_2\rho^{j}_2)|x+\ra+(c_3\rho^{*-j}_1+c_4\rho^{*-j}_2)|x-\ra.
\end{equation}
This result is further constrained by imposing the boundary conditions. In order to find localized wave function near $j=1$, one imposes the boundary conditions
$\phi=(0,0)^T$, such that the general solution  now writes
\begin{equation}
\psi_j=c_+(\rho^{j}_1-\rho^{j}_2)|x+\ra+c_{-}(\rho^{*-j}_1-\rho^{*-j}_2)|x-\ra.
\end{equation}
For an edge state, one cannot have $|\rho|=1$ since this would correspond to a freely propagating mode perpendicular to the edge. The localized solutions at $j=1$ exist if and only if one of two following conditions is satisfied:
\begin{equation}\label{cond1}
|\rho_1|>1,\qquad|\rho_2|>1,\qquad c_+=0,
\end{equation}
or
\begin{equation}\label{cond2}
|\rho_1|<1,\qquad|\rho_2|<1,\qquad c_-=0.
\end{equation}

Thus if there is an edge state solution, its form and energy are
\begin{align}
\psi_j^{(+)}&=c_+(\rho^{j}_1-\rho^{j}_2)|x+\ra\notag\\
E&=2t_1\cos(k_x)
\end{align}
or 
\begin{align}
\psi_j^{(-)}&=c_{-}(\rho^{*-j}_1-\rho^{*-j}_2)|x-\ra\notag\\
E&=-2t_1\cos(k_x).
\end{align}
where
\begin{equation}
\rho_{1,2}=\frac{-b\pm\sqrt{b^2-4ac}}{2a},
\end{equation}
with
\begin{eqnarray}
a&=&-it_1+t_2e^{ik_x}-it_3,\notag\\
b&=&m+2t_3\sin(k_x),\notag\\
c&=&-it_1+t_2e^{-ik_x}+it_3.
\end{eqnarray}

In summary, there can be at most two solutions localized at $j=1$. The existence conditions (\ref{cond1}) and (\ref{cond2}) are imposed to determine the number of edge states and their extension. Note that for every solution at edge $j=1$ there is the solution at $j=L_y$ which can be readily obtained by inversion and conjugation of $\rho$'s.

The topological properties of the bulk phase manifest themselves as existence of edge states at $j=1$. When $\text{Ch}=0$ no edge state solution exists, since one of $|\rho_i|$ is larger then one and the other is smaller than one. When $|\text{Ch}|=1$, only one of the solutions holds in the Brillouin zone. When 
$|\text{Ch}|=2$, there are $k_x$ where both solutions hold.

Let us illustrate the above results for a special point $t_1=1, t_2=1, t_3=0, m=0$ of the phase diagram (\ref{fig:PhDiag}). This point corresponds to the ``center'' of the ${\rm Ch}=-2$ phase and is characterized by the largest gap and flattest bands for the spectrum of the bulk states. One can expect two edge states at either end of the cylinder. The eigenmodes $\rho$ are determined by
\begin{equation}
\rho_{1,2}=\pm i\frac{\sqrt{(-it_1+t_2e^{ik_x})(-i t_1+t_2e^{-ik_x})}}
{-it_1+t_2e^{ik_x}}.
\end{equation} 

Therefore
\begin{equation}
|\rho_{1,2}|=\bigg(
\frac{t_1^2+t_2^2+2t_1t_2\sin(k_x)}{t_1^2+t_2^2-2t_1t_2\sin(k_x)}\bigg)^{1/4},
\end{equation}
and the localization length $\xi$ of the edge state is given by
\begin{equation}
\xi_{1,2}=-1/\ln(|\rho_{1,2}|).
\end{equation}

It is apparent, either by plotting the eigenmodes' absolute value or by checking the existence condition of Eq.~(\ref{ECond}), that edge states are expected for any $k_x$. Therefore this choice of parameters proves to be quite particular, giving maximal extension for edge states.

Evaluation of $|\rho|$ shows that $|\rho|\geq 1$
for $k\in [0,\pi]$ and   $|\rho|\leq 1$ for
$k\in [-\pi,0]$. That means the edge state localized near $j=1$ (bottom) is an eigenstate $|x-\ra$ in the BZ interval $(0,\pi)$ and the other is an eigenstate of $|x+\ra$ in $(-\pi,0)$. The $j=L_y$ solution (top) immediately follows by symmetry.

\begin{figure}[t]
\centering
\includegraphics[width=\columnwidth]{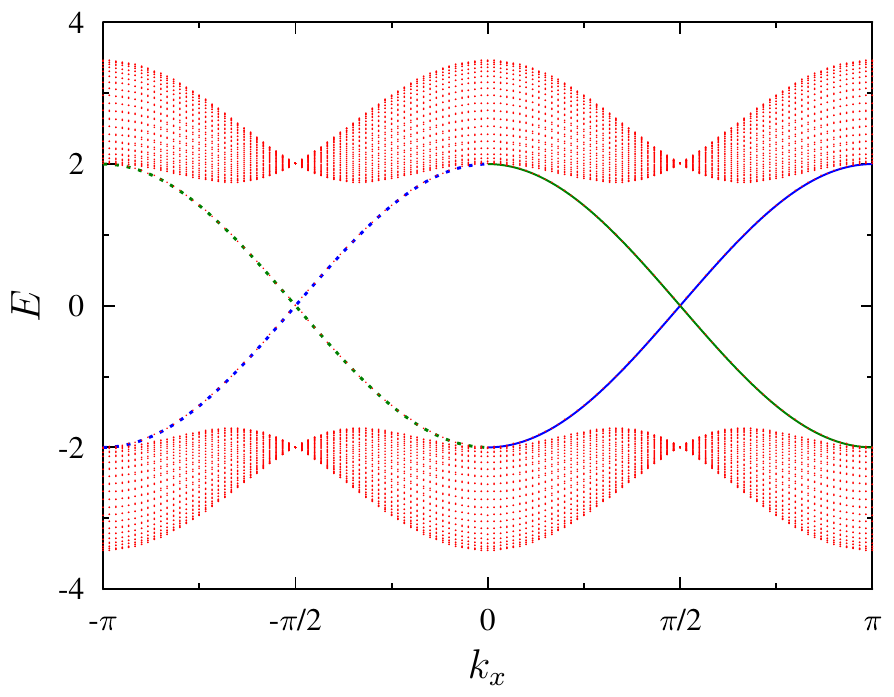}
\caption{Energy spectrum as a function of momentum $k_x$ on a cylindrical geometry (height 40 sites and circumference 180 sites). Two edge states located near $j=1$ are represented in blue, and two at $j=40$ in green. The position and chirality of the edges is schematically represented in the inset. (color online)}
\label{fig:edge}
\end{figure}

As expected, there are four edge states, with wave functions and energies given by: 
\begin{align}\label{origsol}
\psi_j^{-Kb}&=c_+(1-(-1)^j)\rho_1^j|x+\ra\notag\\
E&=2t_1\cos(k_x),&k_x&\in(-\pi,0)\notag\\
\psi_j^{Kb}&=c_-(1-(-1)^j)\rho^{*-j}_1|x-\ra\notag\\
E&=-2t_1\cos(k_x)&k_x&\in(0,\pi)\notag\\
\psi_j^{Kt}&=c_+(1-(-1)^j)\rho_1^j|x+\ra\notag\\
E&=2t_1\cos(k_x),&k_x&\in(0,\pi)\notag\\
\psi_j^{-Kt}&=c_-(1-(-1)^j)\rho_1^{*-j}|x-\ra\notag\\
E&=-2t_1\cos(k_x),&k_x&\in(-\pi,0).
\end{align}
The indices $t$ and $b$ indicate whether the edge states live close to the top ($j=L_y$) or the bottom ($j=1$) part of the cylinder. $\pm K$ indicates whether the edges state crosses the zero energy at $\pm\pi/2$ or, equivalently in this case, whether it is extended in the right, respectively left, part of the BZ. The coefficients $c_{\pm}$ are normalization coefficients which are not of interest here. The edge states' spectra (\ref{origsol}) are plotted in Fig.~(\ref{fig:edge}) together with the numerical solution.

Notice that the pair of states living at either end of the cylinder have the same direction of propagation. Taking the derivative of their dispersion shows that the velocity is identical. Still they are distinguished by a ``valley'' quantum number $\pm K$, determined by where they cross zero energy. 

\section{\texorpdfstring{$\mathbb{Z}_2$}{Z2} insulator}
\label{sec:Z2}
In this section, a quantum spin Hall ($\mathbb{Z}_2$) insulator is created out of the Chern insulator in Eq.~(\ref{BH}) by adding a spin flavor to the model. In particular, we show that the edge states
constructed from the $\pm 2$ Chern numbers are not robust and actually
correspond to a trivial insulator, in agreement with the $\mathbb Z_2$
classification. This is confirmed by the existence of one-particle, TRI, local perturbations that open the spectral gap.

We suppose that the third component of the spin $s_3$ commutes with the Hamiltonian. Thus the system consists of two independent components. We suppose that the spin up component is described by (\ref{BH}), while the spin down one represents its time reversed copy. This yields the following 4-band Hamiltonian:
\begin{eqnarray}\label{4BH}
H(\mb k) &=&2t_1\cos(k_x)\sigma_1\otimes s_0+2t_1\cos(k_y)\sigma_2\otimes s_3\notag\\
&&+[m+2t_2\cos(k_x+k_y)]\sigma_3\otimes s_0\notag\\
&&+2t_3(\sin(k_x)+\sin(k_y))\sigma_3\otimes s_3,
\end{eqnarray} 
where $s$ represents electronic spin and $\sigma_0$, $s_0$ are identity matrices.

Because the spin Hamiltonian is created by two copies of the spinless Hamiltonian, with no spin mixing terms, the conditions for the energy gap are not changed. That means the previously found insulating phases remain insulating phases for the new model. 

Edge states are again found when going to the cylindrical geometry explored in the previous section. Because there are no spin mixing terms, the energy spectrum is  trivially obtained by ``doubling'' the spectra already found for the spinless Hamiltonian. More precisely it is obtained from the union of the spinless (now spin up) Hamiltonian spectrum and its reflection about $k_x=0$ under TRS. Therefore the number of edge states will also double such that each original edge state will get its Kramers partner.

Although every previously nontrivial Chern phase will show edge states in the spinful model, not all of them are robust. Indeed, as we shall see, the QSH insulator constructed out of the spinless model with ${\rm Ch}=\pm 2$ allows for a one-particle TRI perturbation that destroys the edge states.\cite{Wu, Koenig}

The low-lying edge states are described by an effective Hamiltonian, obtained by linearizing the solutions (\ref{origsol}) near $k_x=\pm\pi/2$. At a given edge, for the phase with ${\rm Ch}=\pm 2$, this yields:
\begin{align}\label{effedge}
H_{\rm eff}(q_x)&=
\Psi^\dag_{KR\up}vq_x\Psi_{KR\up}+\Psi^\dag_{-KR\up}vq_x\Psi_{-KR\up}\notag\\
&\quad-\Psi^\dag_{-KL\down}vq_x\Psi_{-KL\down}-\Psi^\dag_{KL\down}vq_x\Psi_{KL\down},
\end{align}
where Fermi velocity reads $v=2t_1$. The indices of the fermionic creation and annihilation operators $\Psi^\dag$ and $\Psi$ describe the valley ($\pm K$), the direction of motion ($L$ or $R$) and that of spin ($\up$ or $\down$). Note also that the first two terms in $H_{\rm eff}$ describe the dynamics of spin up electrons, and therefore correspond to the original 2-band Hamiltonian, while the spin down terms stem from of time reversal operator $T$; $T\Psi_{KR\up}T^{-1}=\Psi_{-KL\down}$. The locking between the direction of the spin and that of motion means that $H_{\rm eff}$ describes a helical liquid.\cite{Wu} 

The edge states above are not robust because one can create the following one-particle, TRI, local perturbation that will gap the edge helical liquid in 
Eq.~(\ref{effedge}) (local meaning there is no inter-edge scattering)
\begin{equation}
\Psi^\dag_{KR\up}\Psi_{KL\down}-\Psi^\dag_{-KR\up}\Psi_{-KL\down}+{\rm h.c.}.
\end{equation}

It is possible to build many tight-binding perturbations yielding the above form at low energy. It is noteworthy to observe that they all break the spin $s_3$ symmetry. An example of TRS perturbation in the tight-binding formulation is
$t_4\sin(k_x)\sigma_3\otimes s_1$.

For the phases with ${\rm Ch}=\pm 1$ no one-particle, local, TRI perturbation can result in backscattering of the edge states. The above example agrees with the statement that models with an even number of Kramers pairs of edge states are $\mathbb Z_2$-trivial.\cite{KMZ2}

\section{Conclusion}
\label{sec:sum}
We have developed a  way to calculate the Chern number by a discrete summation for 2-band, two-dimensional, TRS breaking, topological insulators in a tight-binding formulation. The method allows for an efficient determination of topological phases in such systems and helps to conceive models with large Chern numbers without increasing the number of energy bands. 
We start with simple gapless Hamiltonian containing only two Pauli matrices and $2n$ Dirac points.  Then the system is gapped by adding an explicitly designed mass term containing the third Pauli matrix, yielding any of $2n+1$ possible topological phases
(the Chern number can thus vary between $\{-n,-n+1,\dots,n\}$).

This technique is illustrated by constructing an ``artificial'' Chern insulator which has five available topological insulating phases described by the Chern numbers $\{0,\pm 1,\pm 2\}$. After determining completely the phase diagram and the bulk properties of the model, we have studied the edge structure of the model on a finite cylindrical geometry. For the particular choice of edge geometry, the conditions for existence of gapless edge states and analytical expressions for their dispersion were obtained. The edge wave function structure is also determined for the edge states in any topologically nontrivial phase. For the case $\text{Ch}=-2$, we consider in detail the edge state solution and show it to be in agreement with numerical simulations. Finally we discuss a 4-band realization of a QSH $\mathbb{Z}_2$-insulator from the 2-band Chern insulator at hand. We determine the phase diagram of the 4-band model and show that the Chern phases $\pm 2$ indeed yield the edge states unstable with respect to one-particle TRI perturbations.

The technique described in the present paper could be used to design 2-band tight-binding models featuring the QAH effect with
large Chern number, realizable with ultracold fermionic atoms
in an optical lattice. Recently, several 2D optical lattices---honeycomb,
brick-wall, kagome, checkerboard, etc.---with more than one orbital per
unit cell were realized experimentally.\cite{Jo, Tarruell, Dalibard} In addition, methods to create artificial gauge potentials--and therefore complex hopping
amplitudes---are now reaching maturity.\cite{Dalibard} Thus it appears feasible to tailor an optical lattice that could feature non-trivial Chern insulator. The result could be verified experimentally by measuring the Berry curvature through Bloch oscillations, as recently
shown,\cite{Tarruell, Price} or time-of-flight measurements.\cite{Zhao, *Alba}

\begin{acknowledgments}
The authors would like to thank M.O. Goerbig and G. Montambaux for interesting discussions.
\end{acknowledgments}
\bibliography{bibl}
\end{document}